\documentclass{moriond}

\def\Journal#1#2#3#4{{#1} {\bf #2}, #3 (#4)}

\def\PRL{\em Phys. Rev. Lett.}
\def\PRD{{\em Phys. Rev.} D}
\def\JHEP{{\em JHEP} }

\def\be{\begin{equation}}
\def\ee{\end{equation}}
\def\bea{\begin{eqnarray}}
\def\eea{\end{eqnarray}}

\usepackage{amsmath}
\usepackage{amsfonts}
\usepackage{amssymb}
\usepackage{mathtools}
\usepackage{braket}

\def\gr#1{{\color{green} #1}}
\def\rd#1{{\color{red} #1}}

\usepackage{graphicx}

\usepackage{color}
\definecolor{nicered}{rgb}{0.7,0.1,0.1}
\definecolor{nicegreen}{rgb}{0.1,0.5,.1}

\usepackage{float}

\usepackage{comment}

\newcommand{\Photo}{\includegraphics[height=35mm]{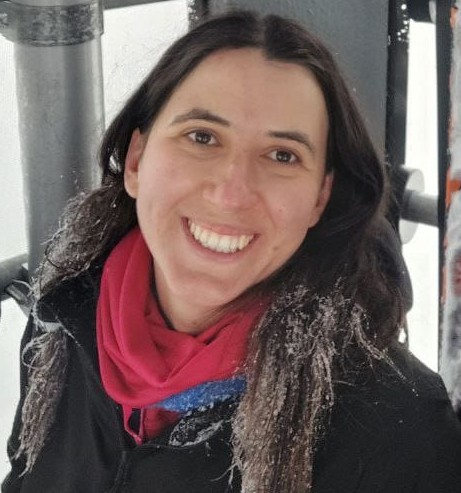}}

\begin{document}
\vspace*{4cm}
\title{Resonances in $D^0\to\pi^+\pi^-\ell^+\ell^-$ and sensitivity to New Physics}

\author{ Eleftheria Solomonidi }

\address{Departament de Física Teòrica, Instituto de Física Corpuscular, \\
Universitat de Val\`encia—Consejo
Superior de Investigaciones Científicas, \\
Parc Científic, Catedrático José Beltrán 2,
E-46980 Paterna, Valencia, Spain}

\maketitle\abstracts{
The study of processes involving the weak decays of the charm quark offers unique possibilities to test the Standard Model from the up-type sector and to probe different New Physics scenarios. In here we focus on the rare decays $D^0\to\pi^+\pi^-\ell^+\ell^-$. The effectiveness of the Glashow-Iliopoulos-Maiani mechanism in charm results in a series of observables that serve as null tests of the Standard Model. Simultaneously, the dominant Standard Model contribution to such decays comes from long-distance dynamics. Following previous literature we describe this with the mediation of resonances. Motivated by recent experimental results on analogous semileptonic decays we investigate the effect of the $S$-wave of the pion pair by considering the scalar resonance $f_0(500)$. The experimental mass distributions and angular observables are described well with our model and there is a significant improvement on the agreement when accounting for $f_0(500)$, which amounts to around 20\% of the total branching ratio. Furthermore, we propose and estimate the size of a series of additional observables to be measured, which can help probe the $S$-wave and act as complementary null tests of the Standard Model. }

\section{Introduction}

Processes involving flavour-changing neutral currents (FCNCs) of quarks have long played a central role in flavour phenomenology. Being suppressed by the Glashow-Iliopoulos-Maiani (GIM) mechanism that arises from the unitarity of the CKM mixing matrix, they are suppressed in the Standard Model (SM) and thus make for a promising ground for the discovery of New Physics (NP). An example of an extensively studied rare process is that of the weak decay $b\to s\ell^+\ell^-$, predominantly with $\ell=\mu$, where a lot of the literature has been dedicated to controlling the hadronic uncertainties related to local form factors of the sort $B\to K(^*)$ as well as to non-local dynamics. 

In this proceeding we focus on an analogous but less studied process that is the decay $c\to u\ell^+\ell^-$.  
The charm quark enjoys a special position among quarks, as it is the only one of the up type that is bound in hadrons and decays weakly. Therefore its decays give access to novel combinations of the CKM matrix elements, allow for tests of the SM that are complementary to the ones performed from the bottom or strange sectors, and by extension can help in the investigation of NP scenarios with different couplings to quarks. The main feature of the decays of focus is that the FCNCs are extremely suppressed, because of the effectiveness of the GIM mechanism in the charm case also at the one-loop level, as all the internal quarks are much lighter than the $W$ boson. Consequently, the decays $c\to u\ell^+\ell^-$ are driven almost exclusively by long-distance dynamics, which are theoretically challenging to describe as they are driven by non-perturbative QCD.

%\begin{figure}
%    \centering
%    \includegraphics[width=0.5\textwidth]{fcncwithzgamma.png}
%    \caption{An FCNC diagram of the charm quark.}
%    \label{fig:FCNCdiagram}
%\end{figure}

In the formalism of a low-energy effective theory around the mass of the charm, the effective interaction Hamiltonian that describes the decays of interest is: 

\begin{equation} \label{eq:H_eff}
	\mathcal{H}_{\rm eff} = \frac{G_F}{\sqrt{2}} \left[ \, \sum^2_{i = 1} C_i (\mu) \left( \lambda_d Q_i^d + \lambda_s Q_i^s \right) - \lambda_b \left( C_7 (\mu) Q_7 + C_9 (\mu) Q_9 + C_{10} (\mu) Q_{10} \right) \right] + \mathrm{h.c.}
\end{equation}
where \begin{eqnarray}\label{eq:operator_list}
	&& Q_1^{d_i} = ( \overline{d_i} c )_{V-A} ( \overline{u} d_i )_{V-A} \,, \\
	&& Q_2^{d_i} = ( \overline{u} c )_{V-A} ( \overline{d_i} d_i )_{V-A} \,, \quad d_i=d,s, \, \nonumber\\
    && Q_7 = \frac{e}{8 \pi^2} m_c \overline{u} \sigma_{\mu \nu} (\mathbf{1} + \gamma_5) F^{\mu \nu} c \,, \nonumber\\
    && Q_9 = \frac{\alpha_{em}}{2 \pi} ( \overline{u} \gamma_\mu (\mathbf{1} - \gamma_5) c ) ( \overline{\ell} \gamma^\mu \ell ) \,, \nonumber\\
    && Q_{10} = \frac{\alpha_{em}}{2 \pi} ( \overline{u} \gamma_\mu (\mathbf{1} - \gamma_5) c ) ( \overline{\ell} \gamma^\mu \gamma_5 \ell ) \,. \nonumber
\end{eqnarray}
The effectiveness of the GIM mechanism results in the Wilson coefficients (WCs) of the operators 7 and 9 being very small, while the one of $Q_{10}$ is zero up to the considered order in $G_F$ and electromagnetic interactions. Therefore the decay $c\to u\ell\ell$ is driven by long-distance insertions of the current-current operators $Q_1^{d_i}$ and $Q_2^{d_i}$ in combination with electromagnetic interactions. 

In here we examine the four-body decays $D^0\to\pi^+\pi^-\mu^+\mu^-$, which have been recently analysed by LHCb. \cite{LHCbbrs,LHCbangular} They can be described in terms of five kinematical variables, namely two invariant masses and three angles. Because of the limited available phase space of those decays, depicted in Fig.~\ref{fig:phaseSpace}, the differential decay rates with respect to the pion-pair or the muon-pair invariant masses, $p^2$ and $q^2$ respectively, are overwhelmingly populated by resonance peaks. This feature makes the detection of any NP effect, resonant or not, impracticable in the decay-rate distributions. However, due to the high multiplicity of the final state, a large number of angular observables can be constructed by integrating over the angular variables in various combinations. Moreover, due to the approximate absence of $Q_{7-10}$ a lot of those angular observables vanish in the SM and thus constitute null tests of it.\cite{Hiller,Cappiello}

\section{Framework}

\begin{figure}
    \begin{minipage}{0.48\textwidth}
    \centering
    \includegraphics[width=0.9\linewidth]{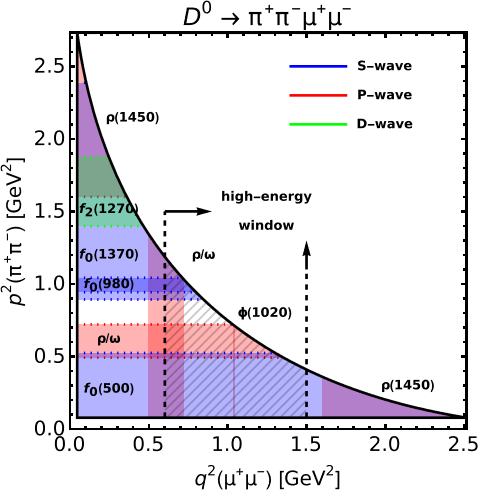}
    \caption{Allowed phase space for the studied decay.}
    \label{fig:phaseSpace}
    \end{minipage}
    \hspace{0.02\textwidth}
\begin{minipage}{0.5\textwidth}
    \centering
    %\includegraphics[width=0.44\textwidth]{W_type_diagram.png}
    %\hspace{3mm}
    %\includegraphics[width=0.44\textwidth]{J_type_diagram.png}
    \includegraphics[width=0.7\linewidth]{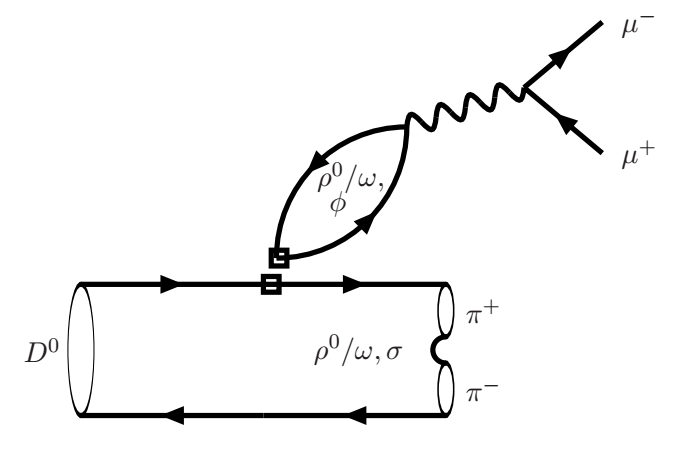}
    \includegraphics[width=0.7\linewidth]{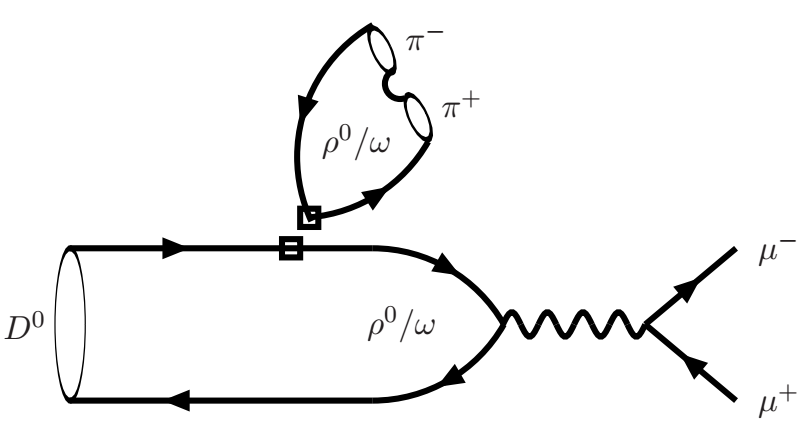}
    \caption{Topologies contributing to the decays of interest in the naive factorisation approach, named W- (top) and J-type (bottom).}
    \label{fig:topo}
\end{minipage}
\end{figure}

In our work,\cite{ourpaper} we focus on the energy region indicated in Fig.~\ref{fig:phaseSpace} as ``high-energy window", in which the process can be adequately described by quasi-two-body decays: the decay of the $D$ meson to two resonances $D^0\to \mathcal{RV}$, out of which the resonance $\mathcal{R}$ decays strongly to two pions while $\mathcal{V}=\rho^0(770)\equiv \rho^0$ or $\omega$ or $\phi$ is a vector resonance that decays electromagnetically to a photon then leading to the muon pair. While previous approaches \cite{Prelovsek,Feldman} have considered the $\mathcal{R}$ to be only the vector resonance $\rho^0$ with a small, isospin-violating $\omega$ admixture, the novelty of our work lies in including the scalar resonance $f_0(500)$, also known as $\sigma$. The presence of such a resonance has been experimentally observed in the analogous semileptonic decays, where it has been measured to account for around 25\% of the total branching ratio.\cite{BES,BES2}

We assume factorisation to hold for the decay $D^0\to\mathcal{RV}$, or equivalently assume that Zweig's rule is in effect. Thus only the neutral-current operator $Q_2^{d_i}$ with the appropriate down-type flavour $d_i$ contributes to the decays. There are three possible topologies that lead to the final state. The naive-factorisation amplitude corresponding to the topology where the initial meson annihilates approximately vanishes. The other two topologies are depicted in Fig.~\ref{fig:topo} and are respectively named W- and J-type. For $\mathcal{R}=\rho$ or $\omega$ and simultaneously $\mathcal{V}=\rho$ or $\omega$ the J-type contraction is as sizeable as the W-type, while the latter results in a $D\to\mathcal{\pi\pi}$ matrix element similarly to semileptonic decays. The presence of the J-type contraction prevents expressing the decay amplitudes in terms of an effective $C_9(q^2)$ coefficient, although in practice one can assign an approximate such coefficient separately for the $S$-wave and the $P$-wave of the pion pair, $C_9^{\mathrm{eff:} S}(q^2)$ and $C_9^{\mathrm{eff:} P}(q^2)$.

The inclusion of further QCD dynamics beyond naive factorisation in the hadronic description is realised in two ways. Predominantly it is achieved with the implementation of line shapes for the intermediate resonances: for the $\rho^0$ decaying to the two pions the well-known Gounaris-Sakurai parameterisation is used, adding a relativistic Breit-Wigner for the $\omega$. The $\sigma$ being a wide resonance is implemented using the Bugg parameterisation\cite{bugg} which takes into account pion scattering data and was also used by BESIII.\cite{BES} Secondly, we assign to each $D^0\to\mathcal{RV}$ vertex an additional strong phase $\delta_{\{\mathcal{R},\mathcal{V}\}}$, assumed to be approximated by a constant, and a ``fudge factor" $B_{\mathcal{V}}^{(S)/(P)}$, allowing for the modification of the magnitudes of the individual amplitudes. 

Our first goal is to give an adequate SM description of the two differential branching ratios over the dipion and dimuon invariant masses with the model described above. The branching ratios are sensitive to some of the free parameters, in particular to the differences between strong phases of the same partial wave as well as to the ``fudge factors". We then fit those parameters to the experimental data. With the fitted values for the parameters, we provide predictions for the angular observables. Given that the SM sources of CP violation can be neglected for the current level of precision, we consider the CP-antisymmetric angular observables to vanish. We then compare our angular observables to the CP-symmetric measured observables $S_i$.

The angular observables exhibit different interference patterns, both between different partial waves and between SM and NP, which are summarised in Table~\ref{tab:angobs} in terms of the approximate WCs $C_9^{\mathrm{eff:} S}$ and $C_9^{\mathrm{eff:} P}$. There are two possible meaningful ways of integrating over the angle of the pions $\theta_\pi$,
\begin{equation}
    \langle I_i \rangle_{-} \equiv \left[ \int^{+1}_{0} d \cos \theta_\pi - \int^{0}_{-1} d \cos \theta_\pi \right] I_i \,, \quad \langle I_i \rangle_{+} \equiv \int^{+1}_{-1} d \cos \theta_\pi I_i \,.
\end{equation}
We examine the angular observables that can be constructed by both ways of integration for all angular observables. This is to be contrasted to the latest LHCb analysis, which focused only on the observables that do not require the presence of an $S$-wave.

The null-test observables vanish in the SM because of the absence of a SM-induced $C_{10}$ and are given schematically in the table for the case of non-vanishing $C_{10}$ coming from NP. We estimate the size of the null-test observables for the value of $C_{10}$ that saturates the bounds from \cite{fajferKosnik}, while the rest of $C_9^{NP},\, C_9'^{NP}$ and $C_{10}'^{NP}$ are set to zero. 

\begin{table}[]
 \begin{tabular}{|lcc|c|}
        \hline
        \multicolumn{4}{|c|}{$ \langle I_i \rangle_{+} $}  \\
        \hline
        $i$ & $S$-wave & Null test & WCs  \\
        \hline
        $ 1^\dagger $ & \gr{$ \circ $} & & $ | C_9^{{\rm eff}: S} |^2 $, $ | C_9^{{\rm eff}: P} |^2 $    \\
        \hline
        $ 2^\dagger $ & \gr{$ \circ $} & & $ | C_9^{{\rm eff}: S} |^2 $, $ | C_9^{{\rm eff}: P} |^2 $    \\
        \hline
        $ 3^\dagger $ &  \rd{$ \times $} & & $ | C_9^{{\rm eff}: P} |^2 $  \\
        \hline
        $4$ & \gr{$ \checkmark $} & & $ C_9^{{\rm eff}: S} \, (C_9^{{\rm eff}: P})^\ast $    \\
        \hline
        $5$ & \gr{$ \checkmark $} & yes &{\color{red}$ C^{{\rm eff}: S}_9 \, C_{10}^\ast + C_{10} \, (C^{{\rm eff}: P}_9)^\ast $}   \\
        \hline
        $ 6^\dagger $ &  \rd{$ \times $} & yes & {\color{red}$ \text{Re} \left[ C^{{\rm eff}: P}_9 \, C_{10}^\ast \right]$}  \\
        \hline
        $7$ & \gr{$ \checkmark $} & yes & {\color{red}$ C^{{\rm eff}: S}_9 \, C_{10}^\ast + C_{10} \, (C^{{\rm eff}: P}_9)^\ast $}  \\
        \hline
        $8$ & \gr{$ \checkmark $} & & $ C_9^{{\rm eff}: S} \, (C_9^{{\rm eff}: P})^\ast $  \\
        \hline
        $ 9^\dagger $ &  \rd{$ \times $} & & $ | C_9^{{\rm eff}: P} |^2 $  \\
        \hline
    \end{tabular}
\quad
    \begin{tabular}{|lcc|c|}
        \hline
        \multicolumn{4}{|c|}{$ \langle I_i \rangle_{-} $}  \\
         \hline
        $i$ & $S$-wave & Null test & WCs \\
        \hline
       $1$ & \gr{$ \checkmark $} & & $ C_9^{{\rm eff}: S} \, (C_9^{{\rm eff}: P})^\ast $   \\
        \hline
        $2$ & \gr{$ \checkmark $} & & $ C_9^{{\rm eff}: S} \, (C_9^{{\rm eff}: P})^\ast $   \\

        \hline
        $4^\dagger$ &  \rd{$ \times $} & & $ | C_9^{{\rm eff}: P} |^2 $  \\
        \hline
        $5^\dagger$ &  \rd{$ \times $} & yes &  {\color{red}$ \text{Re} \left[ C^{{\rm eff}: P}_9 \, C_{10}^\ast \right] $}  \\
        \hline
        $7^\dagger$ &  \rd{$ \times $} & yes &  \color{red}{$ \text{Re} \left[ C^{{\rm eff}: P}_9 \, C_{10}^\ast \right] $}  \\
        \hline
        $8^\dagger$ &  \rd{$ \times $} & & $ | C_9^{{\rm eff}: P} |^2 $ \\
        \hline
    \end{tabular}
    \caption{Angular observables. The ones with a dagger have already been measured by LHCb. The ones with a circle receive an $S$-wave contribution additionally to the $P$-wave, while the ones with a checkmark are only sensitive to the $S$- and $P$-wave interference. The observables serving as null tests are also shown. The ``WCs" column shows the dependence on the WCs, where $C_{10}$ only comes from NP.}
    \label{tab:angobs}
\end{table}

\section{Results}

\begin{figure}
    \centering
    \includegraphics[width=0.45\textwidth]{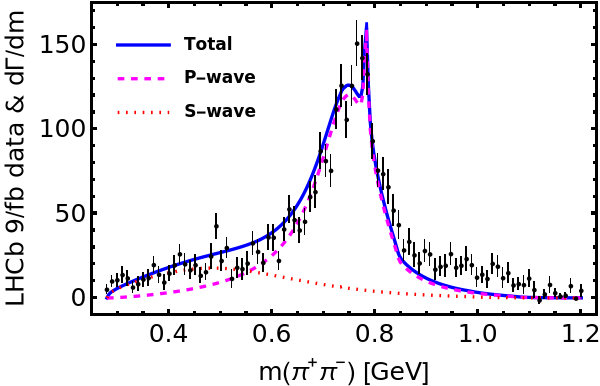}
    \hspace{3mm}
    \includegraphics[width=0.45\textwidth]{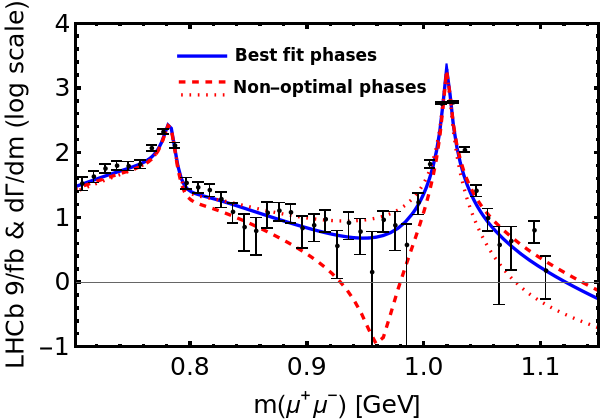}
    \caption{The predictions and the experimental data for the differential decay rate over
the dihadron (left) and the dilepton (right) invariant mass. The normalisation of the vertical axis corresponds to the
number of LHCb events (in logarithmic scale for the right panel).}
    \label{fig:dBrs}
\end{figure}

The theoretical predictions for the two differential branching ratios over $\sqrt{p^2}$ and $\sqrt{q^2}$ result in an overall good agreement with the experimental data, which can be visually verified in Fig.~\ref{fig:dBrs}. The differential distribution over the mass of the pion pair serves for constraining the $\mathcal{R}$-related parameters. Namely, the inclusion of the $\sigma$, which does not manifest as a narrow-resonance shape, results in a significant improvement of the fit in the low-$\sqrt{p^2}$ region. The fit also prefers a very suppressed coupling of $D\to\sigma\phi$ i.e. a small $B_{\phi}^{(S)}$, which if larger would result in a different shape of the $p^2$ distribution; the suppression of such decay mode is in agreement with the amplitude analysis of the analogous four-body hadronic decays.\cite{AmplitudeAnal}

There is a visible sharp $\omega$ peak overlapping with the broader $\rho^0$ peak. The fit constrains the relative size and phase of the $\omega$ admixture with respect to $\rho^0$. The values obtained are compatible with those extracted from semileptonic decays, although the presence of the J-type topology renders meaningless a direct comparison between the two decays even in the absence of further rescattering effects. Finally, a discrepancy between the data and our curve is clearly seen at a dipion mass of around 1 GeV, which can be attributed to various unaccounted-for effects such as the $f_0(980)$ resonance and the opening of the $KK$ channel. This dipion-mass region is not expected to affect the dimuon-mass-dependent observables predicted for the considered $q^2$-window, as indicated in Fig.~\ref{fig:phaseSpace}. 

The differential branching ratio over the dimuon invariant mass exhibits generally good agreement with the data. There is a $\rho$ shape partially overlapping with the sharper $\omega$ peak, which is not suppressed as in the previous case of the strongly decaying resonances (as isospin-symmetry arguments do not apply to electromagnetic decays of the resonances). Note that the presence of the two competing topologies W and J results in a suppressed $D\to{\rho^0(\to\pi\pi)\omega(\to\mu\mu)}$ contribution; therefore the $\omega$ peak comes predominantly from $D\to\sigma\omega$, thus also serving to constrain the $S$-wave. The $\phi$ peak is clearly pronounced on the right of the plot. The fit probes the relative phases $\delta_{ \{ \rho^0/\omega, \rho^0 \} }-\delta_{ \{ \rho^0/\omega, \phi \} }$ and $\delta_{ \{ \sigma, \rho^0 \} }-\delta_{ \{ \sigma, \omega \} }$ from the inter-resonant regions. Given that different phases result in predictions for the $dBr$'s right of the $\phi$ peak which can differ by orders of magnitude (see the red dashed and dotted curves on the right plot of Fig.~\ref{fig:dBrs}), probing those phases is important for future searches of NP in the high-energy region, as the experimental sensitivity will be improved. Finally, we extract from \cite{LHCbbrs} an overall normalisation of around 1.8 accompanied with a high uncertainty (partly due to the presence of a poorly known $D\to\rho^0$ form factor at zero $q^2$), indicating the general aptness of the factorisation approach.

With the fitted parameters, we estimate the values of the SM-dominated observables $S_2,\, S_3$ and $S_4$ to be in good agreement with data in the $q^2$-bins of our interest. The agreement in the first of these observables is significantly improved with the inclusion of the $S$-wave, while the other two are unaffected by the $S$-wave by construction. The observables $S_8$ and $S_9$ are identically zero in our framework, even in the presence of NP. There is some tension with the measurements of those observables in some $q^2$-bins, which could be alleviated with the inclusion of additional strong dynamics, in the form of extra phases related to different transversity form factors, extra resonances or higher partial waves. 

We then turn to the $S$-wave-sensitive observables which are not yet measured by LHCb. Given that these depend on a relative phase between the $S$- and $P$-wave $\delta_{ \{ \sigma, \rho^0 \} }-\delta_{ \{ \rho^0/\omega, \rho^0 \} }$, which is not accessed by the latest experimental analysis, we estimate those observables in terms of the unprobed phase. We find values somewhat smaller than the other, purely $P$-wave angular observables, but still sizeable and within the current experimental sensitivity. We also suggest the measurement of the differential decay rate over the dipion angle, as it is expected to exhibit a clear asymmetry in the presence of the $S$-wave. Its measurement, which can also be performed by integrating over different slices of the dipion mass (see also \cite{babar}), will help determine the so far unprobed phase mentioned and discover further $S$-wave structures. 

Finally we estimate the effect of NP in terms of a non-vanishing $C_{10}$, saturating current experimental bounds. We find the contribution of such NP to the branching ratios and the SM-dominated observables to be negligible. We calculate the null-test observables that depend both on the $S$- and on the $P$-wave interference and find them to be comparable in size to each other and of the order of a few percent. The latest measurements of the null tests do not indicate the presence of NP and the current level of both theoretical and experimental precision does not allow the interpretation of the measured null tests in terms of improved bounds on NP-induced WCs. CP violation is not observed in these decays (for non-leptonic decays see \cite{usToni}).

\section{Conclusions}

We have studied the rare decays $D^0\to\pi^+\pi^-\ell^+\ell^-$, for which we have calculated the SM prediction and examined the effect of potential heavy-scale NP. Due to the effectiveness of the GIM mechanism and the limited phase space, the only sizeable SM contribution to the amplitude comes from long-distance physics and in particular through the mediation of resonances. In our approach we consider the effect of resonances in quasi-two-body topologies, for which we use factorisation assisted with appropriate resonant line shapes, and we accommodate further rescattering effects in constant phases and normalisation factors that are fitted to data. We include for the first time an $S$-wave component of the amplitudes in the form of the $\sigma$ resonance. We find good agreement with the experiment with regard to the differential distributions over the dipion and the dimuon invariant mass and the measured angular observables, and observe a significant improvement resulting from the inclusion of the $S$-wave. We also propose the measurement of and provide predictions for a series of $S$-wave-sensitive, SM-dominated observables, which will help probe so far unaccessed parameters.

We have also provided predictions for the null tests of the SM, in the presence of NP. The current experimental measurements of those observables are compatible with the SM value of zero. However the future experimental sensitivity is expected to be improved down to the level of our predictions for $C_{10}$ at its current-bound value, which is of the order of a few percent. We also stress that the $S$-wave- and $P$-wave-sensitive observables are predicted to be of the same order, and that the $S$-wave null tests will be better predicted theoretically once the $S$-wave SM observables are measured. 

In conclusion, we have managed to gain an overall good control over the SM background, which renders feasible setting meaningful bounds on NP-induced WCs in the next experimental analysis. If a few discrepancies in the SM-dominated observables persist, further refinement of the hadronic model might be needed.

\section*{Acknowledgments}

I would like to thank the organisers for the invitation and the opportunity to present my work in such an exciting conference. Thanks to Luiz Vale Silva for, as always, carefully reading this proceeding and providing his instructive comments. This work has been supported by MCIN/AEI/10.13039/501100011033, Grant No. PRE2018-085325 and by Generalitat Valenciana, Grant No. PROMETEO/
2021/071.

\end{document}